\documentclass[lettersize,journal]{IEEEtran} 
\usepackage{amsmath,amsfonts}
\usepackage{algorithmic}
\usepackage{algorithm}
\usepackage{array}
\usepackage{siunitx}
\usepackage[caption=false,font=normalsize,labelfont=sf,textfont=sf]{subfig}
\usepackage{textcomp}
\usepackage{stfloats}
\usepackage{url}
\usepackage{chemformula}
\usepackage{verbatim}
\usepackage{graphicx, xcolor}
\usepackage{cite}

\begin{document}

\title{Bistability in coupled opto-thermal micro-oscillators}

\author{Aditya Bhaskar, Mark Walth, Richard H. Rand, Alan T. Zehnder

\thanks{This material is based upon work supported by the National Science Foundation under Grant No. CMMI-1634664. The experimental work was performed in part at the Cornell NanoScale Facility, a member of the National Nanotechnology Coordinated Infrastructure (NNCI), which is supported by the National Science Foundation (Grant NNCI-2025233). The numerical work used the Extreme Science and Engineering Discovery Environment (XSEDE), which is supported by National Science Foundation grant number ACI-1548562. Specifically, it used the Bridges-2 system, which is supported by NSF award number ACI-1928147, at the Pittsburgh Supercomputing Center (PSC). This work also made use of the Cornell Center for Materials Research Shared Facilities which are supported through the NSF MRSEC program (DMR-1719875).}

\thanks{A. Bhaskar is with the Sibley School of Mechanical and Aerospace Engineering, Cornell University, Ithaca, NY 14853 USA. Email: {\tt\small ab2823@cornell.edu}.}

\thanks{M. Walth is with the Department of Mathematics, Cornell University, Ithaca, NY 14853 USA. Email: {\tt\small msw283@cornell.edu}.}

\thanks{R. H. Rand is with the Sibley School of Mechanical and Aerospace Engineering and the Department of Mathematics, Cornell University, Ithaca, NY 14853 USA. Email: {\tt\small rhr2@cornell.edu}.}

\thanks{A. T. Zehnder is with the Sibley School of Mechanical and Aerospace Engineering, Cornell University, Ithaca, NY 14853 USA. Email: {\tt\small atz2@cornell.edu}.}

}



\maketitle

\begin{abstract}
In this work, we experimentally investigate the dynamics of pairs of opto-thermally driven, mechanically coupled, doubly clamped, silicon micromechanical oscillators, and numerically investigate the dynamics of the corresponding lumped-parameter model. Coupled limit cycle oscillators exhibit striking nonlinear dynamics and bifurcations in response to variations in system parameters. We show that the input laser power influences the frequency detuning between two non-identical oscillators. As the laser power is varied, different regimes of oscillations such as the synchronized state, the drift state, and the quasi-periodic state are mapped at minimal and high coupling strengths. For non-identical oscillators, coexistence of two states, the synchronized state and the quasi-periodic state, is demonstrated at high coupling and high laser power. Experimentally, this bistability manifests as irregular oscillations as the system rapidly switches between the two states due to the system’s sensitive dependence on initial conditions in the presence of noise. We provide a qualitative comparison of the experimental and numerical results to elucidate the behavior of the system. 
\end{abstract}


\section{Introduction} \label{sec:Introduction}

\IEEEPARstart{M}{icro-} and nano-electromechanical systems (MEMS and NEMS) provide a rich testing ground for studying nonlinear phenomena. Flexural MEMS devices exhibit nonlinearities that can be mechanical in nature resulting from geometric effects such as large deformations, or 
may arise from the devices' interactions with the external environment such as thermal modulation of stress, nonlinear radiation pressure, electrostatic or magnetomechanical forces etc. A survey of the origins of nonlinearities in MEMS devices is given in \cite{tiwari2019using}. Nonlinear effects in the sub-micron scale can be beneficial and have been utilized to create novel MEMS devices such as gyroscopes, energy harvesters, filters, and stable time-keeping oscillators \cite{nitzan2015self, taheri2017operation, ando2010nonlinear, alastalo2005intermodulation, antonio2012frequency}. An active line of research exploits nonlinearities in MEMS devices to study dynamical phenomena \cite{agrawal2013observation, matheny2019exotic, rodrigues2021optomechanical}. The short time scales, scope for innovative device design using established microfabrication techniques, and fine control of the system parameters are advantageous for an experimental study of nonlinear dynamics. In the present work, we use pairs of coupled MEMS oscillators to study different oscillation regimes associated with different system parameters.  

We refer to oscillators that draw energy from a steady source and maintain oscillations via an interchange with a dissipation mechanism, as \textit{Limit Cycle Oscillators (LCO)}. The term is inspired by dynamical systems such as the foundational Van der Pol oscillator which are described by a stable limit cycle in the phase plane \cite{strogatz2000nonlinear}. In this paper, LCOs are also simply referred to as \textit{oscillators}. LCOs, in contrast to resonators, do not require an external periodic forcing function to maintain steady oscillations. In the literature, they are variously referred to as self-sustained oscillators, active oscillators, or autonomous oscillators \cite{jenkins2013self}. Mathematical models for coupled, non-identical LCOs show phenomena such as synchronization to a common locking frequency in the presence of coupling forces, entrainment of oscillators by an external sinusoidal drive, and synchronization in the presence of noise \cite{pikovsky2001universal}. When the LCOs are nearly identical and weakly coupled their dynamics can be studied through their phase evolution alone \cite{strogatz2000kuramoto}. In the present work, we avoid this assumption and use numerical techniques to study the full amplitude-phase equations and discuss the various oscillation regimes such as the drift state, the synchronized state, and the quasi-periodic state, and highlight the coexistence of two stable states of oscillations in the system, also referred to as \textit{bistability}. Previous work on the Van der Pol oscillator system reveals that for a linearly coupled pair of oscillators, stable in-phase and out-of-phase synchronization states can coexist \cite{chakraborty1988transition, osipov2007synchronization}. For a system with nonlinear coupling, regimes of multistability have been demonstrated with a multi-frequency attractor and a chaotic attractor coexisting \cite{kengne2014regular}. Third-order models for LCOs with simple linear coupling also support coexisting modes of vibration \cite{ shayak2020coexisting}.

\begin{figure}
\centering
\includegraphics[width=0.45\textwidth]{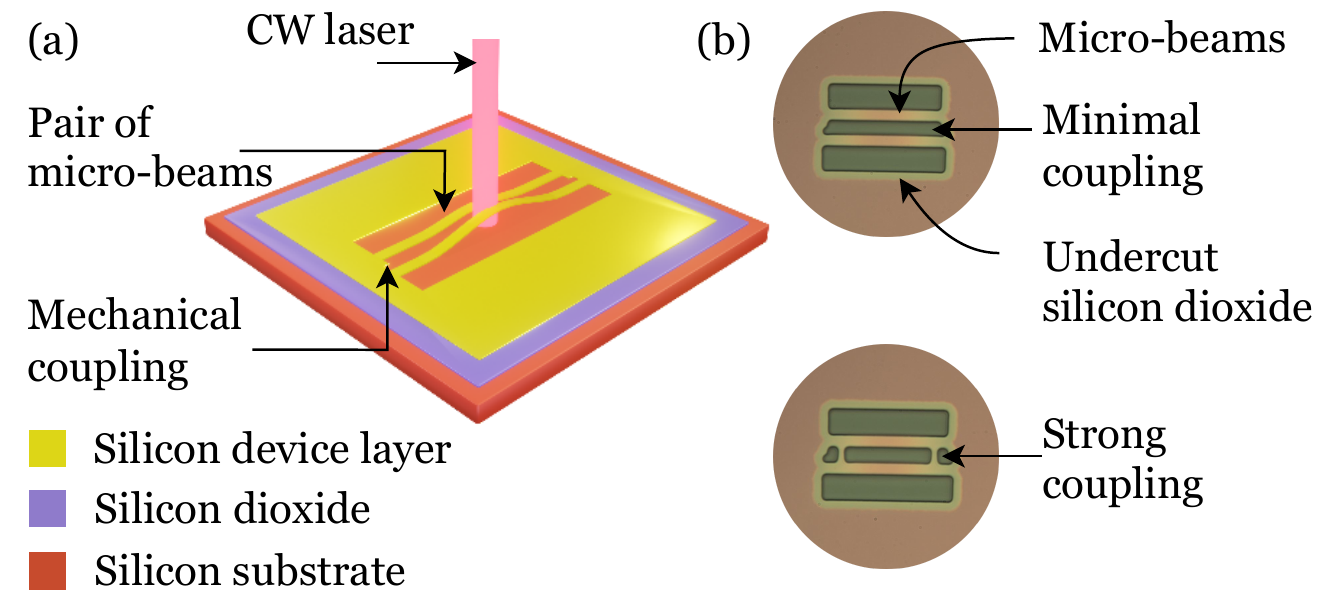}
\caption{(a) Schematic of a pair of mechanically coupled, doubly clamped, silicon micro-oscillators on an SOI chip. The continuous-wave (CW) laser beam drives the silicon beams into out-of-plane limit cycle oscillations. (b) Optical microscope images of silicon beams with minimal mechanical coupling (top), and strong mechanical coupling via bridges (bottom). The undercut silicon dioxide layer is seen as the yellow region surrounding the device features.}
\label{fig:MEMS}
\end{figure}

There has been a significant interest in using MEMS devices to study the nonlinear behavior of coupled LCOs. Synchronization in a pair of coupled LCOs has been observed in systems that are piezoelectrically actuated and electronically coupled \cite{ matheny2014phase}, optically actuated and coupled \cite{zhang2012synchronization}, magnetomotively actuated and mechanically coupled \cite{shim2007synchronized}, and electrically actuated and coupled \cite{agrawal2014synchronization}. Uniform rings of eight LCOs have been used to examine different dynamical states such as  weak chimeras, decoupled states, traveling waves, and inhomogeneous synchronized states in the presence of simple linear coupling, beyond the weak coupling limit \cite{matheny2019exotic}. Higher-order frequency locking of an LCO by an external sinusoidal force has also been recently demonstrated \cite{rodrigues2021optomechanical}. Notably, chaotic and irregular oscillations have been observed in opto-mechanical oscillators driven by radiation pressure due to nonlinearities exhibited by optical cavities at high laser powers \cite{carmon2007chaotic, tallur2015non, navarro2017nonlinear}. We contrast this with the present work where we use opto-thermal laser actuation to drive mechanically coupled, doubly clamped silicon structures into limit cycle oscillations. A schematic of the devices is shown in Fig.~\ref{fig:MEMS}(a). We study two types of devices: minimally coupled devices with mechanical coupling acting via the silicon overhang, and strongly coupled devices with coupling bridges near the anchor points as shown in the optical microscope images in Fig.~\ref{fig:MEMS}(b). Opto-thermal excitation of silicon beams into limit cycle oscillations was first described in \cite{langdon1988photoacoustics} and an interpretation from the perspective of nonlinear dynamics was provided in \cite{aubin2004limit}. The laser power threshold at which the limit cycle is born in a Hopf bifurcation was calculated using an analytical perturbation treatment of a lumped-parameter model for the devices. This analysis was extended to the doubly clamped silicon micro-beam design \cite{blocher2012anchor, blocher2013analysis} which we use in the present work.  It was shown that this model for LCOs exhibits synchronization for a coupled network of two or more oscillators, sensitive dependence on initial conditions for certain system parameters, and entrainment by an external sinusoidal drive \cite{zehnder2018locking, bhaskar2021synchronization}. Further, prior work has been done on a simplified system of equations for the opto-thermally driven, coupled MEMS model to explain the bifurcations, and coexistence of stable dynamical states has been shown \cite{rand2018dynamics, rand2020simplified, shayak2020coexisting}.

In this paper, the results are presented in two main sections. Section \ref{sec:Num} describes the lumped-parameter model for a pair of coupled LCOs, the numerical methods used to solve the system, and the results from the simulations. The different regimes of oscillations are mapped in the laser power vs. coupling strength parameter space. Section \ref{sec:ExpObsv} describes the MEMS device, fabrication techniques, the experimental setup and procedures, and the results from the experiments. The different regimes of oscillations from the numerical simulations and from the experiments are qualitatively compared and bistability induced by changing laser power is demonstrated experimentally. The main result is that the laser power can be used to change the effective frequency detuning between the oscillators while keeping the other parameters in the experiment fixed. This allows us to access regions of bistability, where the steady state solution of the system depends sensitively on initial conditions and can either fall into a state of synchronization or into a state of quasi-periodic oscillations. Experimentally, bistability is seen as irregular oscillations due to the system jumping between the two states in the presence of noise.

\section{Numerical analysis of coupled micromechanical LCOs}\label{sec:Num}

This section presents the computational results for the dynamics of pairs of coupled opto-thermally driven MEMS oscillators. The lumped-parameter model for the MEMS oscillators and the numerical methods to study the model are described in detail. Parameter sweeps are used to unravel the various states exhibited by the oscillators. The coexistence of two stable states of oscillations and sensitive dependence on initial conditions at high input laser powers and strong coupling between the oscillators is highlighted.

\subsection{Mathematical model}\label{ssec:Model}
We use a lumped-parameter model for the opto-thermally driven, mechanically coupled MEMS oscillators. This model was introduced in prior work \cite{aubin2004limit, zehnder2018locking}, and is given by Eqs.~(\ref{eq: displ1})-(\ref{eq: temp2}). The out-of-plane deflection of the center of the micro-beam normalized by the wavelength of the laser source, $z(t)$, and the average temperature of the oscillator, $T(t)$, are the time-dependent variables in the model. Each oscillator is modeled by a third order system containing a pair of ordinary differential equations i.e.  Eqs.~(\ref{eq: displ1}) and (\ref{eq: temp1}) for the first oscillator and Eqs.~(\ref{eq: displ2}) and (\ref{eq: temp2}) for the second oscillator. 
\begin{align}
&\ddot{z}_1 + \frac{\dot{z}_1}{Q} + \kappa_1(1+CT_1)z_1 + \beta z_1^3 + \zeta(z_1-z_2) =DT_1, \label{eq: displ1} \\
&\dot{T}_1 = -BT_1 + HP_{\text{laser}}\left(\alpha + \gamma\sin^2(2\pi(z_1-\bar{z}))\right), \label{eq: temp1}\\
&\ddot{z}_2 + \frac{\dot{z}_2}{Q} + \kappa_2(1+CT_2)z_2 + \beta z_2^3 + \zeta(z_2-z_1) =DT_2, \label{eq: displ2} \\
&\dot{T}_2 = -BT_2 + HP_{\text{laser}}\left(\alpha + \gamma\sin^2(2\pi(z_2-\bar{z}))\right). \label{eq: temp2}
\end{align}

The kinematic equations given by Eqs.~(\ref{eq: displ1}) and (\ref{eq: displ2}) describe the variation of the displacement, $z(t)$ and consist of terms corresponding to a damped harmonic oscillator with unit mass, damping coefficient equal to the inverse of the quality factor, $Q$, and linear stiffness, $\kappa$. Micro-resonators typically have a high quality factor which translates to low damping. In this work, the two oscillators have different fixed linear stiffness values, $\kappa_1=1$ in Eq.~(\ref{eq: displ1})  and $\kappa_2=0.81$ in Eq.~(\ref{eq: displ2}), to model the baseline frequency detuning in the system. The stiffness values are chosen such that the ratio of the linear frequencies of the two oscillators match that of the MEMS oscillators studied experimentally and described in Section \ref{ssec:Devices}. The kinematic equations also have a Duffing-like nonlinearity with cubic stiffness, $\beta$. Thermal feedback to the oscillations is modeled via the static displacement per unit temperature with coefficient, $D$, and the linear stiffness per unit temperature with coefficient, $C$.

The thermal equations given by Eqs.~(\ref{eq: temp1}) and (\ref{eq: temp2}) are obtained by modifying Newton's law of cooling with a coefficient of heat transfer, $B$, by adding a laser absorption term to it. The laser absorption, which is a dimensionless ratio between $0$ and $1$, varies with the gap between the center of the device and the substrate underneath and is approximated by a sine-squared function with constants for the minimum absorption, $\alpha$, the contrast in absorption, $\gamma$, and the normalized position of the minimum of the absorption curve relative to the device’s equilibrium position, $\bar{z}$. The coefficient of thermal absorption is given by $H$. The majority of the model parameter values are derived in \cite{blocher2012optically} and reproduced in Table ~\ref{tab:tab1} for completeness. However, the cubic stiffness here is negative to model the amplitude-softening behavior of the oscillators seen in the experiments. Furthermore, the coefficient of linear stiffness per unit temperature is higher than the previously-derived value. The oscillators are driven at a constant laser power, $P_{laser}$, and are symmetrically coupled via a linear term with coupling strength, $\zeta$.
\begin{table*}
\centering
\begin{tabular}{llll}
\hline\noalign{\smallskip}
Parameter & Symbol & Value & Units  \\
\noalign{\smallskip}\hline\noalign{\smallskip}
Quality factor & $Q$ & $1240$ & [AU] \\
Cubic stiffness & $\beta$ & $-10$ & [AU] \\
Coefficient of static displacement per unit temperature & $D$ & $2.84 \times 10^{-3}$ & \si{1/K} \\
Coefficient of linear stiffness per unit temperature & $C$ & $0.04$ & \si{1/K} \\
Coefficient of heat transfer & $B$ & $0.112$ & [AU] \\
Minimum absorption & $\alpha$ & $0.035$ & [AU] \\
Contrast in absorption & $\gamma$ & $0.011$ & [AU]  \\
Minimum of absorption w.r.t device equilibrium & $\bar{z}$ & $0.18$ & [AU] \\
Coefficient of thermal absorption  & $H$ & $6780$ & \si{K/W} \\
\noalign{\smallskip}\hline{\smallskip}
\end{tabular}
\caption{Fixed parameter values used in Eqs.~(\ref{eq: displ1})-(\ref{eq: temp2}) for the numerical study. [AU] stands for arbitrary units.} 
\label{tab:tab1} 
\end{table*}
\subsection{Numerical methods} \label{ssec:NumMethods}
The model given by Eqs.~(\ref{eq: displ1})-(\ref{eq: temp2}) was numerically integrated primarily for the time series, $z(t)$, of each oscillator. The input laser power, $P_{laser}$, the coupling strength, $\zeta$, and the initial displacements of the oscillators, $z_1(0)$ and $z_2(0)$ were varied in the model to study their effects on the dynamics. The other initial conditions were fixed for all runs at $\dot{z}_1(0)=\dot{z}_2(0)=T_1(0)=T_2(0)=0$. Each simulation was run for a time span of $50Q$ [AU] so that any initial transient behavior had decayed completely. The steady-state dynamics of the two oscillators were studied via the resultant frequency spectra, with a frequency resolution of $\approx 1.5\times 10^{-6}$ [AU]. 

The synchronization behavior of the pair of oscillators is classified as \textit{drift state} if the peaks of the spectra corresponding to the two oscillators do not coincide, and satellite peaks i.e. peaks in the spectra different from the limit cycle frequency, are absent. The behavior is classified as \textit{synchronized state} if the peaks of the spectra corresponding to the two oscillators coincide and the satellite peaks are absent. The behavior is classified as \textit{quasi-periodic state} if the peaks of the spectra corresponding to the two oscillators do not coincide and satellite peaks are present. We performed parameter sweeps and analyzed the resultant spectra to classify the states of oscillations for different model parameters. Parallel computing was used to perform the parameter sweeps over $251\times 251$ parameter grids \cite{towns2014xsede}.

\subsection{Numerical results}\label{ssec:NumResults}
\begin{figure*}
\centering
\includegraphics[width=\textwidth]{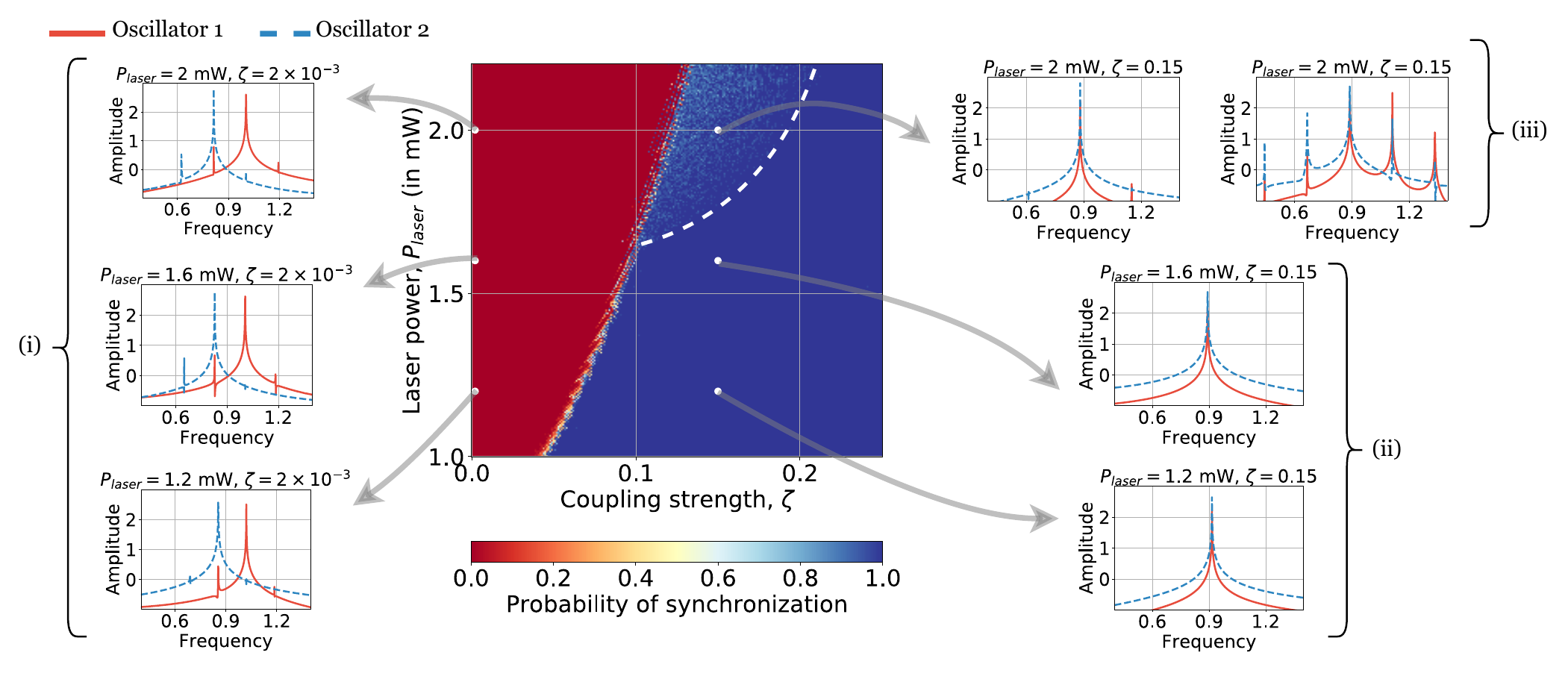}
\caption{Numerical heat map of the probability of synchronization for a pair of coupled micromechanical LCOs at different laser powers and coupling strengths. For sample parameter values, three qualitatively different dynamics are indicated by the corresponding spectra: (i) drift state at minimal coupling strengths ($\zeta = 2\times 10^{-3}$) (ii) stable synchronized state in the presence of strong coupling at low laser powers ($\zeta = 0.15$ and $P_{laser} = 1.2$ or $1.6$ \si{mW}) (iii) bistability i.e. coexistence of stable synchronized oscillations and stable quasi-periodic oscillations in the presence of strong coupling at high laser powers ($\zeta = 0.15$ and $P_{laser} =2$ \si{mW}).}
\label{fig:NumericalResults}
\end{figure*}
The behavior of the coupled MEMS LCO system is visualized in Fig.~\ref{fig:NumericalResults}. The two parameters in the system that are varied are the laser power, $P_{laser} \in [1, 2.2]$ \si{mW}, and the coupling strength, $\zeta \in [0, 0.25]$. The laser power is always kept above the Hopf threshold value to get stable limit cycle oscillations in the system. The Hopf thresholds for the LCOs in the model depend on the stiffness values and are approximately $P_{laser}\approx 1$ \si{mW}. A span of coupling strengths, from minimal to strong, was chosen to illustrate qualitatively different oscillations such as the synchronized, quasi-periodic, and drift states. It should be noted that the units of the input source and coupling strength do not directly correspond exactly to the experimental laser power and coupling strength, and are to be understood only qualitatively. For every pair of parameters, $(\zeta, P_{laser})$, the system of differential equations was numerically solved for 25 different initial conditions $z_1(0) \in [-0.2, 0)$ and $z_2(0) \in [0,0.2)$ and the resultant spectra were used to calculate the probability of synchronization for the coupled oscillators. The resulting heat map for the probability of synchronization is plotted in Fig.~\ref{fig:NumericalResults}. 

For minimal coupling strengths, the probability of synchronization is zero for any input laser power. The spectra corresponding to three different values of laser power, $P_{laser} \in \{1.2, 1.6 ,2\}$ \si{mW}, for minimal coupling strength $\zeta = 2 \times 10^{-3}$, are shown in Fig.~\ref{fig:NumericalResults}(i). All three spectra show two distinct peaks corresponding to the frequencies of the two LCOs. Physically, this implies that the two oscillators exert minimal influence on each other and are free-running LCOs. A key observation is that the difference between the frequencies of the LCOs increases with increasing laser power. This suggests that the laser power can be used to change the effective frequency detuning between the oscillators. 

In Fig.~\ref{fig:NumericalResults}, for high coupling strengths, there is a region of certain synchrony for any initial conditions in the chosen range. The boundary of this region varies with the input laser power such that for a fixed coupling strength, if the laser power is increased, the system may lose synchronization. This is shown numerically in a series of three spectra for coupling strength fixed at a high value, $\zeta = 0.15$. For relatively lower laser powers, $P_{laser} = 1.2$ \si{mW} and  $P_{laser} = 1.6$ \si{mW}, the spectra of the two LCOs always collapse into a single peak corresponding to the frequency of synchronization, as shown in Fig.~\ref{fig:NumericalResults}(ii). When the laser power is increased to a higher value, $P_{laser} = 2$ \si{mW}, there are two states of oscillations that the system might exhibit depending on the initial conditions. There is a stable, synchronized state with the two oscillators having coincident frequencies corresponding to the peaks of the spectra, and there is a stable quasi-periodic state where multiple prominent satellite peaks appear in the spectra of the two oscillators. These two stable states are shown in Fig.~\ref{fig:NumericalResults}(iii). Physically, the quasi-periodic state corresponds to a significant influence of each oscillator on the other but failure to synchronize. The coexistence of two stable states is attributed to the increasing frequency detuning with laser power, such that the coupling strength is unable to guarantee synchronization. This can also be inferred from the boundary for certain synchronization in Fig.~\ref{fig:NumericalResults}, where the coupling strength required for certain synchronization increases with increasing laser power. 

\section{Experimental analysis of coupled micromechanical LCOs} \label{sec:ExpObsv}
This section presents the results from the experiments performed on mechanically coupled, opto-thermally driven pairs of MEMS LCOs to reveal their dynamics. The design and fabrication of the MEMS devices are described in brief. The experimental setup to opto-thermally drive and detect the oscillations are also described. Devices in the minimally coupled and in the strongly coupled regimes are analyzed and their dynamics at various laser powers are mapped and discussed.
\subsection{Device design and fabrication} \label{ssec:Devices}
\begin{figure}
\centering
\includegraphics[width=0.35\textwidth]{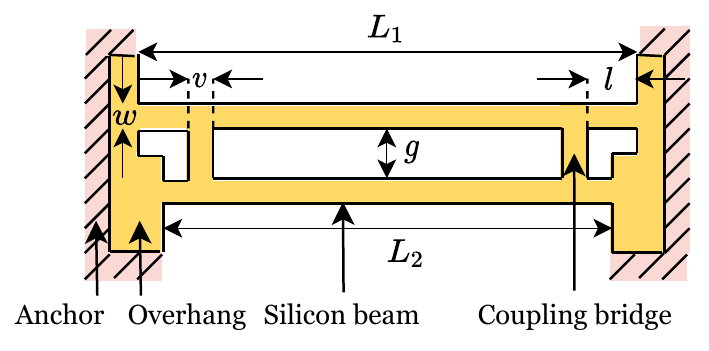}
\caption{Schematic of mechanically coupled, doubly clamped beams with key dimensions labelled (not to scale). For the strongly coupled devices studied in this work, $L_1 = 40$ \si{ \mu m}$, L_2 = 38$ \si{\mu m}$, w =2$ \si{\mu m}$, g = 3$ \si{\mu m}$, l = 3$ \si{\mu m}, and $v = 1$ \si{\mu m}. The thickness of the silicon device layer is $205$ \si{nm}. For minimally coupled devices, the coupling bridges are absent and all other dimensions remain the same.}
\label{fig:SchematicDraw}
\end{figure}
Doubly clamped, beam microresonators were fabricated on a $0.5$ \si{in} $\times$ $0.75$ \si{in} silicon-on-insulator (SOI) chip with a stack consisting of a $205$ \si{nm} silicon device layer on a $400$ \si{nm} silicon dioxide layer with a thick silicon substrate underneath. Single microbeams and coupled pairs of microbeams of various lateral dimensions were fabricated on the chip, out of which devices of certain dimensions were chosen for the final experiments. The dimensions of the strongly coupled pair of microbeams that are investigated in the current paper are labelled in Fig.~\ref{fig:SchematicDraw}. Note that the devices in this schematic are not drawn to scale. In the minimally coupled case, the beams were of the same dimensions but the coupling bridges between the beams were absent. Standard photolithography was used for fabricating the devices. The SOI chip, spin-coated with positive photoresist, was patterned using a $5\times$ g-line stepper with a resolution of $0.9$ \si{\mu m}. After developing the photoresist, the exposed silicon device layer was shallow-etched using inductively coupled plasma reactive ion etching with \ch{C_4F_8}\ch{/SF_6} chemistry. The silicon dioxide layer was then undercut using a $6:1$ buffered oxide etchant to release the devices. The lateral undercut near the anchors was at least $1$ \si{\mu m} which causes the oscillators to exert a minimal influence on each other via the mechanical overhang \cite{sato2006colloquium}. The undercut region can be visually seen in Fig.~\ref{fig:MEMS}(b) as the thick yellow border on the periphery of the device features. Thus, there is always some minimal mechanical coupling between the oscillators even in the absence of explicit coupling bridges. Critical point drying was used after the wet etch release step to circumvent surface tension effects \cite{ carr1998measurement}. Optical profilometry of the released devices revealed that the microbeams are buckled outwards with stronger buckling effect in longer beams. As an example, a pair of micro-beams that were $28$ \si{\mu m} long and $2$ \si{\mu m} wide with coupling bridges between them had a center displacement of about $0.2$ \si{\mu m}. Outward buckling is caused due to the presence of axial compressive stresses in the silicon device layer prior to the release step.  
\subsection{Setup and procedures}\label{ssec:setup}
\begin{figure*}
\centering
\includegraphics[width=0.85\textwidth]{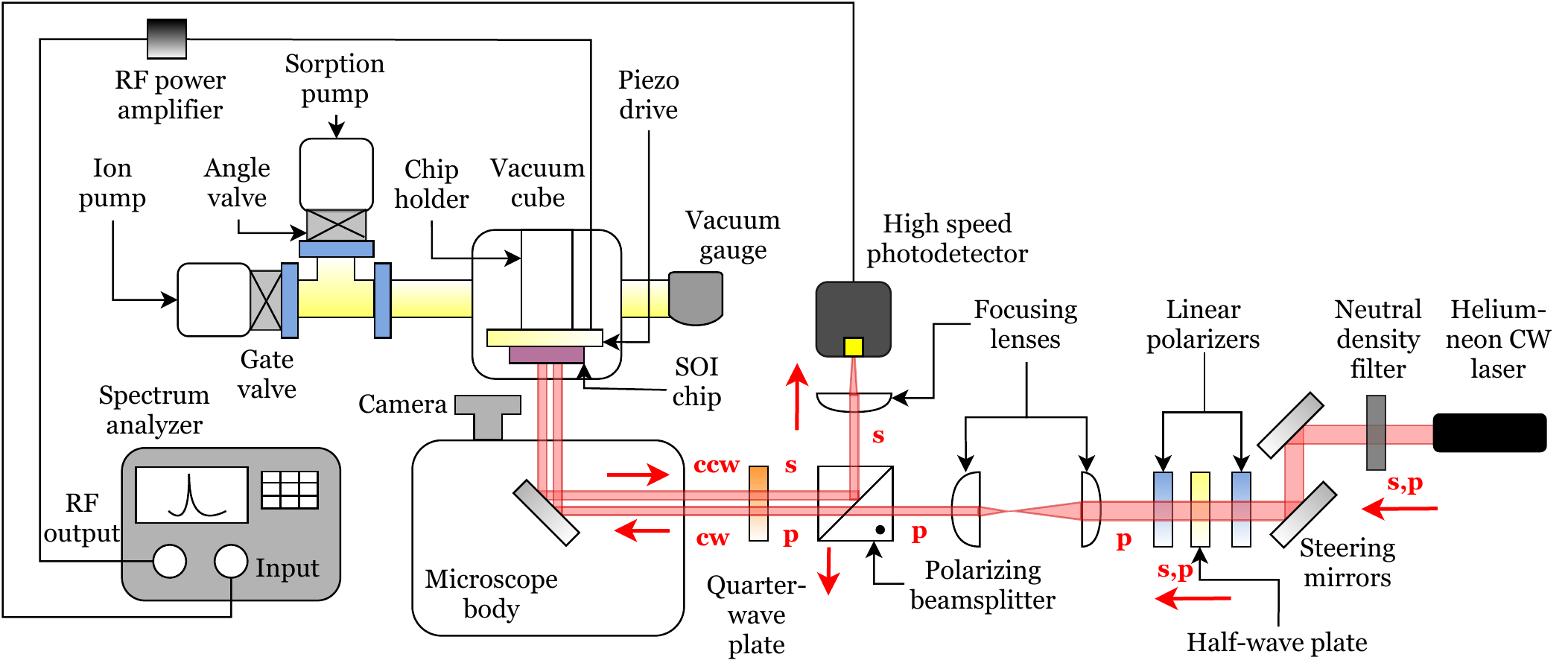}
\caption{Schematic of the experimental setup to analyze the dynamics of coupled MEMS limit cycle oscillators. The SOI chip with the devices is housed in a vacuum cube and opto-thermally driven using a continuous-wave laser. The frequency spectrum of the oscillations is measured through the reflected light collected in the high-speed photodetector. The light reflected from the chip retraces the path of incidence but are shown separately in the schematic for clarity. The piezo-drive is not used in the study of limit cycle oscillations.}
\label{fig:Setup}
\end{figure*}
We now briefly describe the experimental methods used to drive and detect oscillations in the MEMS devices. A schematic of the experimental setup is given in Fig.~\ref{fig:Setup}. The micro-beams were induced to undergo out-of-plane limit cycle oscillations using opto-thermal excitation from a continuous-wave (CW) laser source. In this work we used a $20$ \si{mW} helium-neon CW laser source at a wavelength of $\approx 633$ \si{nm}. The laser beam path was adjusted using a pair of steering mirrors kept in a Z-fold configuration. The power and polarization of the laser beam were controlled  using a half-wave plate sandwiched between two linear polarizers. The transmission axes of the two polarizers were fixed to set the plane of polarization to be parallel to the optical table. The half-wave plate is able to rotate the plane of polarization of the light passing through it without attenuating the intensity. The half-wave plate and the second linear polarizer effectively work as cross-polarizers, and the half-wave plate is rotated to control the laser intensity going into the setup. The laser power going into the setup was calibrated to the rotation angle of the half-wave plate which was recorded while performing the experiments. Focusing lenses were used to reduce the laser spot diameter to illuminate only the devices of interest. The plane of incidence in the current setup is set by the orientation of the polarizing beam-splitter cube and is parallel to the optical table. In Fig.~\ref{fig:Setup}, the polarization of the laser light is marked as p-polarized (parallel to the plane of incidence), s-polarized (perpendicular to the plane of incidence), or circularly polarized: cw (clockwise), or ccw (counterclockwise). The polarizing beam-splitter and a quarter wave plate were used to create an optical isolator such that the light reflected from the devices was redirected to a high-speed photodetector \cite{blocher2012optically}. 

The laser light passes through the optical elements and enters the microscope through a port and is then redirected towards the silicon device layer of the SOI chip. The partially transmissive silicon-device layer, gap in the silicon dioxide layer, and the reflective substrate underneath form a Fabry-Pérot cavity interferometer. The laser light undergoes multiple rounds of transmission, reflection, and absorption, in the cavity such that the net absorbed energy and the net reflected energy are periodic functions of the gap between the center of the silicon device and the substrate underneath \cite{sekaric2003studies}. The absorbed light energy causes thermal expansion of the silicon devices and an out-of-plane deflection. The device dimensions are chosen such that the absorption-displacement curve results in thermal feedback to support stable limit cycle oscillations. This mechanism of thermal feedback is explained in more detail in prior work \cite{blocher2012anchor}. The SOI chip with the devices was indium-bonded to a piezoelectric disc actuator and was held inverted in a vacuum chamber. The piezoelectric actuator is driven using a spectrum analyzer output via a power amplifier to provide external sinusoidal forcing. Note that the piezo-drive is used only in resonance experiments to establish the baseline resonance frequencies of the devices and not in the opto-thermally driven limit cycle oscillation experiments. In the resonance experiments, the laser power is below the Hopf threshold for limit cycle oscillations and the laser is used only for detection of oscillations.

The pressure in the vacuum chamber was reduced in a two-stage process: first using a sorption pump to reduce the pressure to $\approx 10^{-3}$ \si{mBar}, followed by an ion pump to reach a final pressure $< 10^{-6}$ \si{mBar}. The devices are kept in the vacuum chamber to minimize damping and to keep the devices clean. For detection of oscillations, the reflected light modulated by the oscillations of the microbeams was collected in a high-speed photodetector and its frequency components were recorded in a spectrum analyzer. A beam of laser light with a spot diameter greater than the total width spanned by the pair of oscillators was used to both drive and detect the dynamics. The intensities of light reflected from the two oscillators get added and thus the resultant spectrum recorded on the spectrum analyzer was a sum of spectra corresponding to the two oscillators. It should be noted that the actual laser power reaching the device is less than the power entering the microscope due to the losses in the optical path inside the microscope. For a similar setup, this loss was previously measured to be about $65\%$ \cite{blocher2012optically}. The experimental laser powers reported in this work were measured as the laser entered the microscope. Lastly, we mention that the microscope camera and the neutral density filter are two components of the setup that were used only in the laser alignment procedure; the filter to attenuate the laser power, and the camera to visualize the laser spot as we position the laser at the center of the device of interest. Centering the laser spot on the devices is necessary for equal laser power to reach both oscillators. A manual linear stage supporting the entire vacuum chamber was used for centering and for scanning across the chip to study different devices. The laser spot was aligned at the lowest laser power setting such that the intensities of the reflected light from both beams were roughly the same on visual inspection. This protocol was repeated consistently for each experiment.
\begin{figure*}[ht!]
\centering
\includegraphics[width=\textwidth]{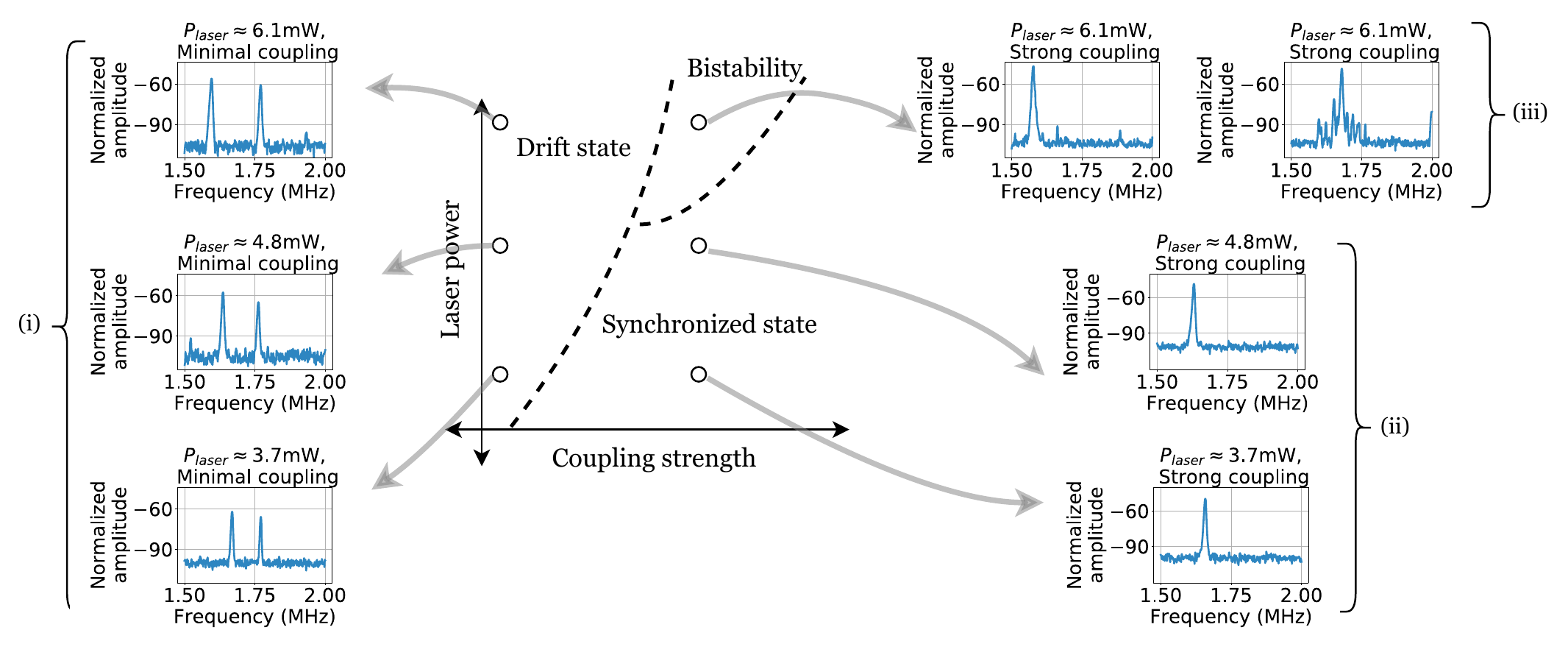}
\caption{A map of the experimentally recorded dynamics of a pair of coupled MEMS oscillators (i) drift state at minimal coupling strengths (ii) synchronized state in the presence of strong coupling at low laser powers (iii) co-existence of two stable states i.e. synchronized oscillations and quasi-periodic oscillations in the presence of strong coupling at high laser powers.}
\label{fig:ExperimentalResults2}
\end{figure*}
\subsection{Experimental results}\label{ssec:Expresults}
We charted the dynamics for two sets of devices at the extremes of our coupling strengths: 1) a pair of oscillators $38$ and $40$ \si{\mu m} long and coupled via bridges with dimensions shown in Fig.~\ref{fig:SchematicDraw}, and 2) a pair of oscillators of the same dimensions minimally coupled via the mechanical overhang, with the coupling bridges absent. The uncoupled resonance frequencies of the micro-beams are roughly $1.655$ and $1.755$ \si{MHz}. We measured the resonance frequencies at laser powers below the Hopf threshold such that the laser is used only for sensing and is not responsible for driving limit cycle oscillations. Base excitation was provided via the piezo-drive.

For the experiments on limit cycle oscillations, the base excitation was turned off and the laser power was always kept at a value higher than the Hopf threshold. The laser power was varied continuously and the two sets of devices, at minimal and strong coupling strengths, were studied. The quantitative coupling strengths are not necessary to distinguish the two regimes. A qualitative map of the different oscillatory responses is shown in Fig.~\ref{fig:ExperimentalResults2}. In the plotted results, the magnitude of measured frequency spectra is normalized by the laser power entering the microscope to consistently represent the relative amplitude of oscillations \cite{blocher2012optically}. At minimal coupling strength, there are two stable, distinct peaks in the measured spectra at the three different laser powers above the Hopf threshold: $3.7, 4.8,$ and $6.1$ \si{mW}. Note that these are the laser power values at the microscope port, not accounting for the losses inside the microscope. The peaks in each spectrum correspond to the drifting LCOs exerting minimal influence on each other and are shown in Fig.~\ref{fig:ExperimentalResults2}(i). It is noted that the difference between the frequencies of the LCOs increases as the laser power is increased. For the strongly coupled case, low laser powers above the Hopf threshold give rise to a stable synchronized state, shown in Fig.~\ref{fig:ExperimentalResults2}(ii). Note that again the frequency of synchronization changes with laser power. 

When the laser power crosses a critical value, a sudden onset of irregular oscillations characterized by a broadband spectrum with fluctuations was observed. Snapshots of the frequency spectra reveal that there are two states that the system exhibits as shown in Fig.~\ref{fig:ExperimentalResults2}(iii): a synchronized state with the two oscillators having identical frequencies, and a quasi-periodic state with prominent satellite peaks in the spectrum suggesting a strong influence of each oscillator on the other. We explain the irregular motion as the system switching between these two states in the presence of noise. The system spends most of the time in the intermediate state between the synchronized state and the quasi-periodic state which is seen as a broadband spectrum. The irregular spectrum is persistent for a few minutes in the experiments, which is long compared  to the time-scale of the oscillator with a period of the order of $1$ \si{\mu s}. Further, the experiment showing irregular oscillations is repeatable with consistent results each time it was performed. Irregular oscillations are not observed when a single oscillator of length $38$ or $40$ \si{\mu m} is driven at high laser powers at around $6.1$ \si{mW}. Such irregular oscillations are also not observed in identical oscillators, minimally coupled or strongly coupled. Further, irregular oscillations are not observed in non-identical oscillators with minimal coupling. This indicates that irregular oscillations observed in the experiments are not chaotic oscillations intrinsic to the optical cavity which have been reported previously \cite{carmon2007chaotic}. Videos of the frequency spectra of the devices as the laser power is swept up and then down are available as supplementary material \cite{2yyx-6669-22}.

\section{Discussion}\label{sec:disc}
\begin{figure}
\centering
\includegraphics[width=0.45\textwidth]{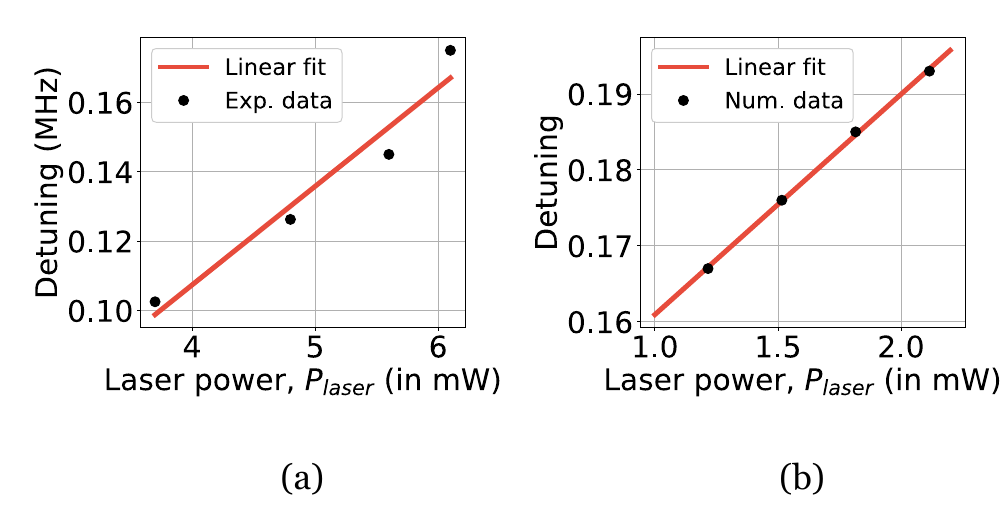}
\caption{Laser-induced frequency detuning between the limit cycle oscillators in the (a) MEMS experiments (b) numerical simulations, with other system parameters held constant.}
\label{fig:DiscussionPlaserFreq}
\end{figure}

Bistability of the synchronized state and the quasi-periodic state is seen in strongly coupled, opto-thermally driven, frequency-detuned oscillators at high laser powers in both the numerical simulations (Section \ref{sec:Num}) and in the MEMS experiments (Section \ref{sec:ExpObsv}). Bistability is precipitated by the increase in effective frequency detuning between the oscillators with the increase in laser power. This effect is shown in Fig.~\ref{fig:DiscussionPlaserFreq}, where the difference in the frequencies between the two oscillators is plotted against the laser power from the numerical and experimental results. It should be noted that the absolute values of the laser power and the frequencies differ between the numerical and experimental results due to the scaling, thus we only capture the dynamics of the system qualitatively. For opto-thermally driven oscillators, the laser power changes the frequency detuning between oscillators. The other parameters that influence the frequency detuning in the system such as the dimensions of the micro-beams, the material, the out-of-plane buckling, are parameters that are either immutable or not smoothly variable. Thus, the laser-induced scheme for frequency detuning can be used to study and access different regimes of coupled oscillator dynamics.

\begin{figure}
\centering
\includegraphics[width=0.35\textwidth]{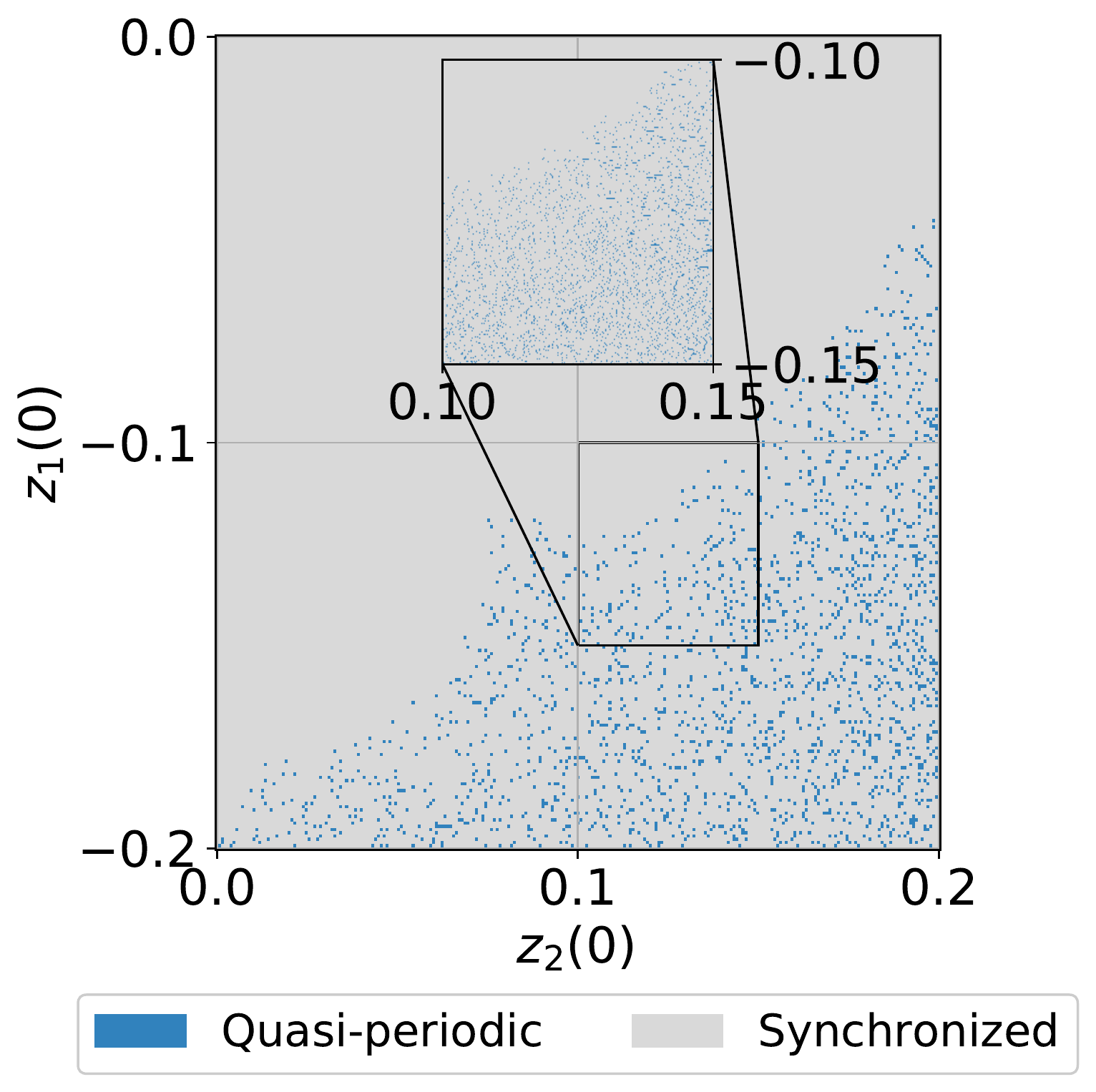}
\caption{Numerical plot showing sensitive dependence of the system on initial conditions $z_1(0) \in [-0.2, 0)$ and $z_2(0) \in [0,0.2)$, at coupling strength $\zeta = 0.15$, and high laser power, $P_{laser}=2$ \si{mW}.} The other initial conditions are fixed at $\dot{z}_1(0)=\dot{z}_2(0)=T_1(0)=T_2(0)=0$. The zoomed inset shows the fine structure in the initial conditions space.
\label{fig:z10vsz20}
\end{figure}

\begin{figure}
\centering
\includegraphics[width=0.45\textwidth]{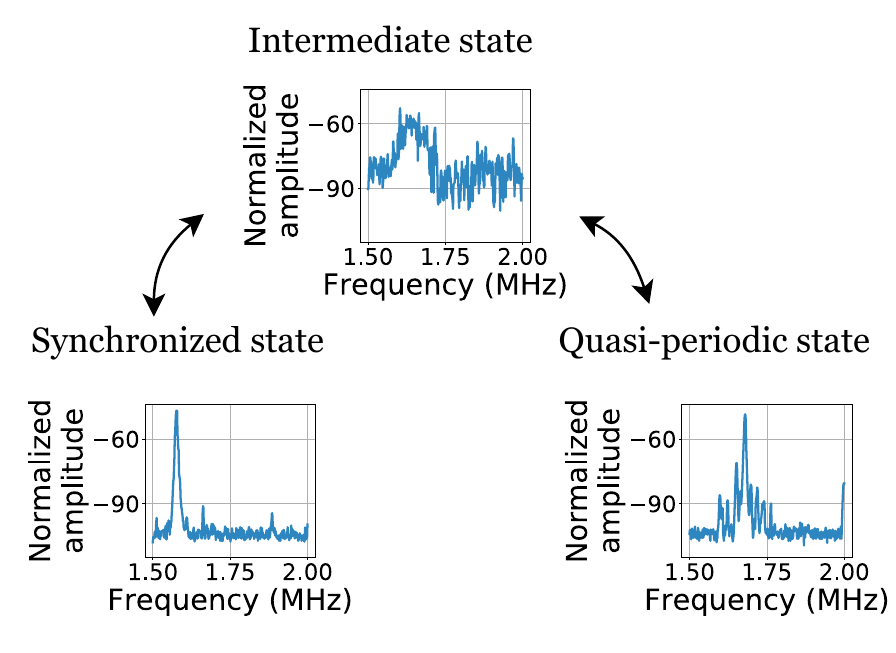}
\caption{Bistability manifests as a broad-band spectral response in strongly coupled oscillators in the experiments (center). The oscillators rapidly switch between the synchronized state (left) with a single prominent peak in the spectrum, and the quasi-periodic state (right) with multiple satellite peaks, due to sensitive dependence of the system on the dynamical variables.}
\label{fig:Irregular}
\end{figure}

Here, we record and discuss the differences between the results in the numerical simulations and the MEMS experiments. The numerical models show two steady-state solutions for the same system parameters at different initial conditions. The basins of attraction for these two states are shown in Fig.~\ref{fig:z10vsz20}, where a $251\times 251$ grid of initial displacements is used. However, no transient behavior is seen in the numerical simulations. This is explained by the fact that there is no noise in the mathematical model. In future work, we plan to investigate the possibility that adding noise to the numerical model can cause the transient behavior that we observe experimentally. The sensitive dependence on initial conditions that is captured in the simulations suggest that any noise present in the experiments such as thermo-mechanical noise or fluctuations in laser power would result in the system switching between the two stable solutions. This is experimentally observed and the intermediate state between the synchronized and quasi-periodic states is a spectrum with a broad-band response as shown in Fig.~\ref{fig:Irregular}. The abrupt onset of irregular spectra associated with the bifurcation to bistability could be exploited in a bifurcation-based sensor application \cite{hajjaj2020linear}.

\section{Conclusions}\label{sec:conc}
In this paper, we have demonstrated the use of laser power to change the frequency detuning between two non-identical, strongly coupled, opto-thermally driven MEMS limit cycle oscillators. This allowed us to access the various regimes of oscillations to study the dynamics of coupled oscillators. A regime of irregular oscillations was experimentally demonstrated and explained using numerical simulations as the system rapidly switching between the synchronized and quasi-periodic states in the presence of noise due to sensitive dependence on initial conditions. This phenomenon was caused by an increase in frequency detuning between strongly coupled oscillators with an increase in laser power. Other states of oscillations such as the drift state, stable synchronized state, and the quasi-periodic state were also experimentally demonstrated and numerically validated. Use of the smoothly variable laser power for changing the frequency detuning in a pair of opto-thermal oscillators is valuable in the study of nonlinear dynamics as well as in potential bifurcation-based sensors. The bifurcation from the stable synchronized state to irregular oscillations is distinct and can be used for the detection of and warning at threshold laser powers. As future work, we expect to carry out a theoretical analysis of the bifurcation to bistability using perturbation theory on a simplified mathematical model for the coupled MEMS system. 

\section*{Acknowledgments}
The authors would like to thank the anonymous reviewers for their suggestions and feedback. 



\bibliographystyle{IEEEtran}

\newpage
 
\vspace{11pt}

\begin{IEEEbiography}[{\includegraphics[width=1in,height=1.25in,clip,keepaspectratio]{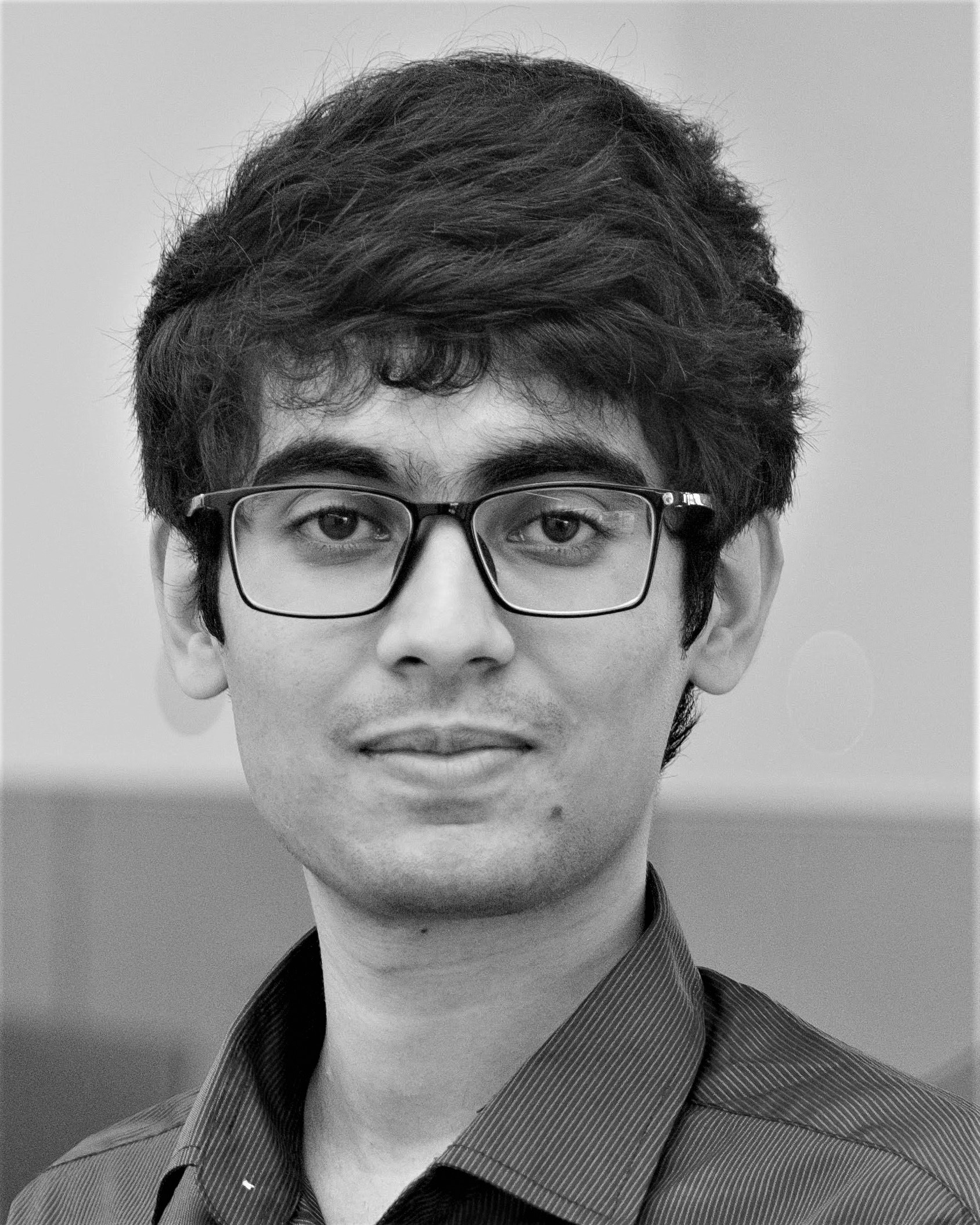}}]{Aditya Bhaskar} was born in Mumbai, India. He received the B.Tech. and M.Tech. degrees in mechanical engineering from the Indian Institute of Technology Madras, Chennai, India, in 2017. He is currently a Ph.D. candidate in the Sibley School of Mechanical and Aerospace Engineering at Cornell University, Ithaca, NY, USA. His research interests include nonlinear dynamics of micromechanical oscillators, distributed averaging algorithms in multi-agent networks, and fracture propagation in heterogeneous materials.    
\end{IEEEbiography}

\begin{IEEEbiography}[{\includegraphics[width=1in,height=1.25in,clip,keepaspectratio]{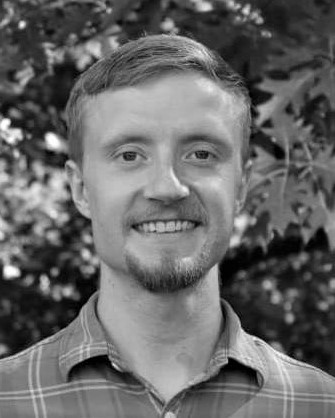}}]{Mark Walth} is a Ph.D. candidate in the Department of Mathematics at Cornell University, Ithaca, NY, working in nonlinear dynamics. Before coming to Cornell, he was a middle school math teacher in Washington, D.C., as a member of the Math for America fellowship program. Mark was born in Colorado Springs, CO, and received his B.A. in mathematics from Reed College in Portland, OR, in 2014. He received his M.A.T. in mathematics education from American University, Washington, D.C., in 2015.   
\end{IEEEbiography}

\begin{IEEEbiography}[{\includegraphics[width=1in,height=1.25in,clip,keepaspectratio]{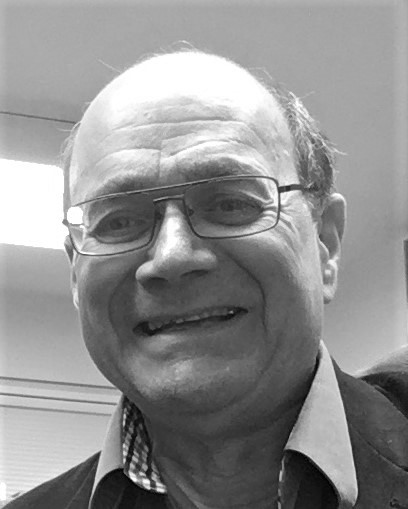}}]{Richard H. Rand} received the B.S. degree from
Cooper Union, New York, NY, USA in 1964, and the M.S. and
Sc.D. degrees from Columbia University, New
York, NY, USA, in 1965 and 1967, respectively, all in civil engineering.
He has been a Professor at Cornell University, Ithaca, NY, USA since 1967 and is currently
in both the Mathematics Department and the Mechanical and Aerospace Engineering Department.
His recent research work involves using
perturbation methods and bifurcation theory to obtain approximate solutions to
differential equations arising from nonlinear dynamics problems in engineering
and biology.
\end{IEEEbiography}

\begin{IEEEbiography}[{\includegraphics[width=1in,height=1.25in,clip,keepaspectratio]{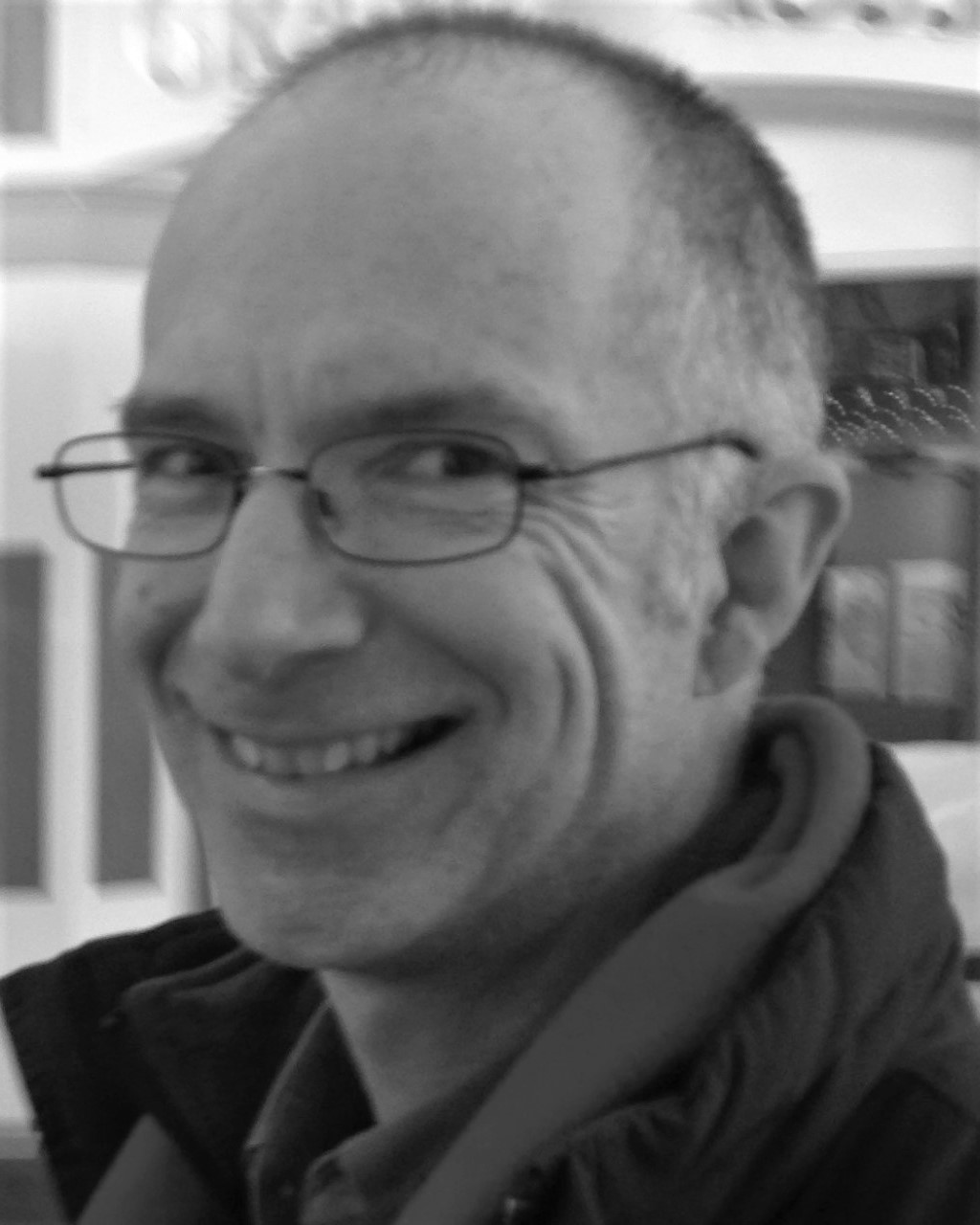}}]{Alan T. Zehnder} received his B.S. degree from the University of California, Berkeley, and Ph.D. from the California Institute of Technology, Pasadena, both in mechanical engineering. He is a Professor in the School of Mechanical and Aerospace Engineering at Cornell University where he serves as the Associate Dean for Undergraduate Programs. His current research interests include the nonlinear dynamics of nanomechanical oscillators and the deformation and fracture of hydrogels. Zehnder is the Editor-in-Chief of \textit{Experimental Mechanics}, a fellow of the American Society of Mechanical Engineers (ASME) and of the Society for Experimental Mechanics (SEM), and a recipient of the S. Nemat Nassar Award from the SEM.
\end{IEEEbiography}

\vfill

\end{document}